
%
\newcount\eqnumber

\def\numbereq{\global\advance\eqnumber by 1 \eqno(\the\eqnumber)}
%
%
\newcount\refnumber

\def\tie{\noexpand~}
\immediate\openout1=refs.tex
\immediate\write1{\noexpand\frenchspacing}
\immediate\write1{\parskip=0pt}
\def\ref#1#2{\global\advance\refnumber by 1%
[\the\refnumber]\xdef#1{\the\refnumber}%
\immediate\write1{\noexpand\item{[#1]}#2}}

%
\def \sqr#1#2{{\vcenter{\vbox{\hrule height.#2pt
	\hbox{\vrule width.#2pt height#1pt \kern#1pt
		\vrule width.#2pt}
		\hrule height.#2pt}}}}
\def\Box{{\mathchoice\sqr65\sqr65\sqr33\sqr23}\,}

\def\mpl#1 {{\it Mod. Phys. Lett.\/ \bf A#1 }}
\def\pl#1 {{\it Phys. Lett.\/ \bf #1B }}
\def\np#1 {{\it Nucl. Phys.\/ \bf B#1 }}
\def\pr#1 {{\it Phys. Rev.\/ \bf D#1 }}
\def\prl#1 {{\it Phys. Rev. Lett.\/ \bf #1 }}
\def\s{(\sigma)}
\def\sp{(\sigma')}
\def\ov{\overline}
\def\Tb{\ov T}
\def\d{\delta(\sigma-\sigma')}
\def\dpr{\delta'(\sigma-\sigma')}

\def\dpppr{\delta'''(\sigma-\sigma')}
\def\ints{\int d\sigma\,}

\def\dx{\partial X}
\def\dbx{{\overline\partial}X}
\def\dsx{\partial^2X}
\def\dbsx{{\overline\partial}^2X}
\def\knl{K^{\nu\lambda}}
\def\half{\textstyle {1\over 2}}
\def\dlinno{\hfill\llap{\global\advance\eqnumber by 1 (\the\eqnumber)}}
\def\eqalinno{{\global\advance\eqnumber by 1}(\the\eqnumber)}
\def\name#1{\xdef#1{\the\eqnumber}}
\font\twelverm=cmr10 scaled\magstep1
\font\eightrm=cmr7 scaled\magstep1
\font\sixrm=cmr5 scaled\magstep1
\font\twelvei=cmmi10 scaled\magstep1
\font\eighti=cmmi7 scaled\magstep1
\font\sixi=cmmi5 scaled\magstep1
\font\twelvesy=cmsy10 scaled\magstep1
\font\eightsy=cmsy7 scaled\magstep1
\font\sixsy=cmsy5 scaled\magstep1
\font\twelvebf=cmbx10 scaled \magstep1
\font\eightbf=cmbx7 scaled \magstep1
\font\sixbf=cmbx5 scaled \magstep1
\font\twelvex=cmex10 scaled \magstep1
\font\twelvesl=cmsl10 scaled \magstep1
\font\twelveit=cmti10 scaled \magstep1
\textfont0=\twelverm \scriptfont0=\eightrm \scriptscriptfont0=\sixrm
\def\rm{\fam0 \twelverm}
\textfont1=\twelvei \scriptfont1=\eighti \scriptscriptfont1=\sixi

\textfont2=\twelvesy \scriptfont2=\eightsy \scriptscriptfont2=\sixsy
\def\cal{\fam2 }
\textfont3=\twelvex \scriptfont3=\twelvex \scriptscriptfont3=\twelvex
\newfam\itfamm \def\it{\fam\itfamm\twelveit} \textfont\itfamm=\twelveit
\newfam\slfamm \def\sl{\fam\slfamm\twelvesl} \textfont\slfamm=\twelvesl
\newfam\bffamm \def\bf{\fam\bffamm\twelvebf} \textfont\bffamm=\twelvebf
\scriptfont\bffamm=\eightbf \scriptscriptfont\bffamm=\sixbf
\headline{\ifnum \pageno>1\it\hfil Deformations, Symmetries and
Topological $\ldots$\/
	\else \hfil\fi}
\footline{\ifnum \pageno>1 \hfil \folio \hfil \else \hfil\fi}
\baselineskip=14pt
\hsize=6in
\hoffset=.24in
\vsize=8.5in
\voffset=.25in
\parindent=3pc
\hfuzz=0.3pt

\rightline{\vbox{\hbox{\rm RU91-14}\hbox{\rm DOE/ER40325-2-TASKB}}}
\vfil
\centerline{\bf DEFORMATIONS, SYMMETRIES AND TOPOLOGICAL}
\bigskip
\centerline{\bf DEGREES OF FREEDOM OF THE STRING\footnote*{\tenrm Talk
presented at the XX Meeting on Differential Geometric Methods in
Theoretical Physics, June 3--7, 1991, at Baruch College, New York City.}}
\vfil
\centerline{\bf Mark Evans and Ioannis Giannakis\footnote{$^{\dag}$}
{\tenrm Work supported in part
by the  Department of Energy Contract Number DOE-ACO2-87ER40325, TASK~B}}
\medskip
\centerline{\it The Rockefeller University}
\centerline{\it 1230 York Avenue}
\centerline{\it New York, NY 10021-6399}
\vfil
\centerline{\bf Abstract}
\medskip
{\narrower
\baselineskip=12pt
\tenrm We discuss three closely related questions; i)~Given a
conformal field theory, how may we deform it? ii)~What are the symmetries
of string theory? and iii)~Does string theory have free
parameters? We show that there is a distinct deformation of the stress
tensor for every solution to the
linearised covariant equations of motion for the massless modes of the
Bosonic string, and use this result to discuss the symmetries of the
string. We also find an additional finite dimensional space of deformations
which may correspond to free parameters of string theory, or alternatively
may be interpreted as topological degrees of freedom, perhaps analogous to
the isolated states found in two dimensions.\bigskip}
\vfil\vfil\break

\line{\bf 1. Introduction \hfil}
\nobreak\bigskip

\rm Conformal field theories (with appropriate central charge) are solutions
to the classical equations of motion of string theory
\ref{\soln}{C.\tie Lovelace, \pl 135 (1984), 75; C.\tie Callan, D.\tie Friedan,
E.\tie Martinec and M.\tie Perry, \np 262 (1985), 593; A.\tie Sen,
\pr 32 (1985), 2102.}, so that by studying infinitesimal deformations
which preserve this conformal structure, we are examining the linearised
classical equation of motion about the corresponding solution. This is
an interesting problem in its own right, but it also gives us insight
into the symmetry structure of string theory. We normally think of
finding symmetries by looking for transformations on the fields which
leave invariant the action of a theory. This
requires that we solve the daunting problems of string field theory
before we can address the problem of symmetry.
However, there is an alternative approach; we may simply work with
the equations of motion of the theory and find transformations which
take one solution into another without changing the physics. A symmetry
is therefore a particular case of a deformation.

Finally, we may shed
some light on the question of whether string theory has free parameters.
For a {\it given\/} conformal field theory, the interactions of all the
physical states appear to be uniquely prescribed, if we want both Lorentz
invariance and unitarity in space-time. Different conformal field theories
are supposed to correspond simply to different background solutions of the
equations of motion. This supposition has a testable consequence, that
every deformation of a conformal field theory correspond to a deformation
of the background value of some physical field of the string. We shall
find that there are in fact deformations of the usual critical bosonic
string which do not correspond to
physical fields, and so may be interpreted as free parameters, or,
equivalently, as isolated states or topological degrees of freedom.
As such they are perhaps higher dimensional analogues of the isolated
states of two-dimensional string theory which have been the focus of
so much discussion at this conference \ref\iso{A. Polyakov, \mpl 6 (1991),
635; B. Lian and G. Zuckerman, \pl 254 (1991), 417 and Yale preprint
YCTP-P18-91; D. Gross, I. Klebanov and M. Newman, \np 350 (1990), 621.}.

\bigbreak
\line{\bf 2. Conformal Field Theory.\hfil}
\nobreak\bigskip

There are several equivalent definitions of a conformal field theory,
of varying degrees of sophistication, but for our purposes we shall take
the simplest which has the added advantage of making the algebraic
properties manifest. In the parlance of physicists we shall consider a
Hamiltonian formulation; that is we shall take our world-sheet to be
a cylinder (twice-punctured sphere if you prefer) with an arbitrarily
chosen cycle around it, parameterised by a single real coordinate
$\sigma$, running from $0$ to $2\pi$. A {\sl Conformal Field Theory\/}
consists of the following, defined on this cycle:
\item{1)} an algebra $\cal A$ of operator valued distributions, usually
called fields
\item{2)} a representation of this algebra, and
\item{3)} two distinguished fields, $T\s$ and $\Tb\s$ which satisfy two
mutually commuting copies of the Virasoro algebra:
\vfil\break

$$
\eqalignno{[T\s, T\sp]&={-i c \over 24\pi}\dpppr
+2i T\sp\dpr - i T'\sp\d&{\global\advance\eqnumber by 1}
(\the\eqnumber a)\name{\eqvir}\cr
[\Tb\s,\Tb\sp]&={i c \over 24\pi}\dpppr
-2i\Tb\sp\dpr + i\Tb'\sp\d&(\the\eqnumber b)\cr
[T\s,\Tb\sp]&=0&(\the\eqnumber c)\cr}
$$
\item{}  and which generate motion around the cycle:

\noindent
$$
\left[L_0-\ov L_0,\phi\sp\right]=
\int d\sigma [T\s-\Tb\s, \phi\sp]=-i\phi'\sp \qquad \forall\; \phi\sp\in
{\cal A}\numbereq\name\eqtrans
$$

\noindent (A prime may denote differentiation with respect to $\sigma$).
These special fields are the two non-vanishing components of the energy
momentum tensor of the field theory, and so include the Hamiltonian, $H$,
$$
H=L_0+\ov L_0=\int d\sigma (T\s + \Tb\s)\numbereq\name\eqham
$$
which may be used to {\it define\/} the evolution of fields off the
cycle, although we shall not use this fact.

Since $Vir \times Vir \subset \cal A$, and any subalgebra acts on its parent
through commutation (the adjoint action), the elements of $\cal A$ will
themselves be grouped into representations of $Vir \times Vir$. It is
therefore natural to define {\sl Primary Fields of dimension $(d,\ov d)$},
which transform simply under the adjoint action, by
$$
\eqalign{[T\s,\Phi_{(d,\overline d)}\sp]&=i d \Phi_{(d,\overline d)}\sp
\dpr-(i/\sqrt2)\partial\Phi_{(d,\overline d)}\sp\d\cr
[\Tb\s,\Phi_{(d,\overline d)}\sp]&=-i\overline d \Phi_{(d,\overline d)}
\sp\dpr-(i/\sqrt2)\overline\partial\Phi_{(d,\overline d)}\sp\d\cr}
\numbereq\name{\eqprim}
$$
The symbols $\partial$ and $\ov\partial$ indicate differentiation with
respect to the light-cone coordinates $x^\pm = (\sigma\pm\tau)/\sqrt2$,
and so take us off the space-like cycle. From our algebraic point of
view, we should think of the symbol $\partial\phi\sp$ as meaning
$i\sqrt2 [L_0,\phi\sp]$ for any field $\phi\sp\in\cal A$, with
a similar meaning for $\ov\partial$. The definition of primary field,
Eq.~\eqprim, is thus an empty tautology for the zero modes of the
energy momentum tensor, but is non-trivial for the others.

To conclude this section, we will try to make clearer what is meant
by the belief that conformal field theories are solutions of the classical
equations of motion of the string. Consider a string moving in some
space-time with metric $G$, then a natural choice for the two-dimensional
field theory to describe this situation is
$$
\eqalign{T\s&=\textstyle{1\over2}G^{\mu\nu}(X)\dx_\mu\dx_\nu\s\cr
\Tb\s&=\textstyle{1\over2}G^{\mu\nu}(X)\dbx_\mu\dbx_\nu\s\cr}
\numbereq\name{\eqcurved}
$$
where the $X$ are scalar fields which can be thought of as coordinates for
the string in space-time, and
$$
\dx_\mu\s={1\over\sqrt2}(\pi_\mu\s+G_{\mu\nu}(X)X^{\nu\prime}\s)\qquad
\dbx_\mu\s={1\over\sqrt2}(\pi_\mu\s-G_{\mu\nu}(X)X^{\nu\prime}\s)
\numbereq\name{\eqd}
$$
and $\pi\s$ is the momentum conjugate to $X$; the only non-vanishing
bracket among the $X$ and $\pi$ is
$$
[\pi_\mu\s, X^\nu\sp] = -i\delta_\mu^\nu\d \numbereq\name\eqpix
$$
(This definition of $\partial X$ is consistent with the one given above).
If our bracket is the Poisson bracket, then the $T$ and $\Tb$ defined
in Eq.~\eqcurved\ satisfy $Vir\times Vir$ for all choices of $G$. However,
if, as we want, they are defined as normal ordered (with respect to the
Fourier modes of $X$ and $\pi$) products of fields, and the bracket
is a commutator, then they satisfy $Vir\times Vir$ only when $G$
satisfies certain conditions which look something like the Einstein
equations of motion. Since the spectrum of the bosonic string includes
a state which has all the properties of a disturbance of the space-time
metric (a, ``graviton,''), this condition of conformal invariance is
naturally interpreted as an equation of motion for this physical field.
It is one of the goals of the work described in this talk to
clarify this relationship between equations of motion for space-time
fields and conformal field theories, and so to generalise it to
include the full, infinite set of space-time fields.

\bigbreak
\line{\bf 3. Deformations of Conformal Field Theories. \hfil}
\nobreak\bigskip

\line{\it 3.1 The Deformation Equations. \hfil}
\nobreak\medskip
Having defined a conformal field theory in the previous section, we
may now consider making an infinitesimal deformation which preserves
the axioms listed above.
We may in principle deform a conformal field theory through any of
its elements, {\it viz.\/}~the algebra $\cal A$, the distinguished
fields $T\s$ and $\Tb\s$ or even the representation (deforming the
cycle should make no difference). However, physicists usually have
a canonical choice for all elements of a theory except its Hamiltonian,
and so we shall consider only changes in the fields $T\s$ and $\Tb\s$.
We are thus interested in deforming the embedding $Vir\times Vir\subset
\cal A$.
The more general problem of deforming a morphism of algebras has
been discussed in the mathematical literature \ref\gerst{M. Gerstenhaber
and S. Schack in {\it Deformation Theory of Algebras and Structures,} ed.
M.\tie Hazewinkel and M.\tie Gerstenhaber (Kluwer, Dordrecht, 1988).}.

We must preserve $Vir \times Vir$, including the value of the central
charge, $c$, and the fact that $L_0-\ov L_0$ generates
translations. This last fact means that $L_0-\ov L_0$ may deform
at most by a central element, and will generally be invariant.
To first order, then, $\delta T\s$ and $\delta\Tb\s$ must
satisfy
$$
\eqalignno{[\delta T\s, T\sp]+[T\s,\delta T\sp]&=2i\delta T\sp\dpr
- i\delta T'\sp\d&{\global\advance\eqnumber by 1}(\the\eqnumber a)
\name{\eqdef}\cr
[\delta\Tb\s,\Tb\sp]+[\Tb\s,\delta\Tb\sp]&=-2i\delta\Tb\sp\dpr
+ i\delta\Tb'\sp\d&(\the\eqnumber b)\cr
[\delta T\s,\Tb\sp]+[T\s,\delta\Tb\sp]&=0&(\the\eqnumber c)\cr}
$$
We shall refer to Eq.~\eqdef\ as the {\sl deformation equations}.
\bigbreak
\line{\it 3.2 Canonical Deformations. \hfil}
\nobreak\medskip

{\sl Let $\Phi_{(1,1)}\s$ be a primary field of dimension (1,1), then the
deformation equations \eqdef\ are satisfied by}
$$
\delta T\s = \delta\Tb\s = \Phi_{(1,1)}\s \numbereq\name\eqcan
$$

This result \ref\dcftss{M.\tie Evans and B.\tie Ovrut, \pr41 (1990), 3149;
\pl231 (1989), 80.} follows from the definition of a primary field,
Eq.~\eqprim, and we shall call such deformations {\sl canonical}. An
alternative, and
world-sheet covariant, discussion of canonical deformations has been
given by Campbell, Nelson and Wong \ref\phil{M.\tie Campbell, P.\tie Nelson
and E.\tie Wong, University of Pennsylvania Report UPR-0439T (1990)}.
Note that such a deformation mixes left and right moving sectors, so
that it is {\it not\/} sufficient to consider just one sector. Also,
since $\delta T\s = \delta\Tb\s$, $T\s - \Tb\s$ is an invariant
of the canonical deformation class, and its zero mode therefore continues
to generate translations, satisfying Eq.~\eqtrans.

Canonical deformations have a number of features which indicate that
they are indeed the way we should, ``turn on,'' a space-time field:
\item{1)} They agree with our preconceptions of how massless fields
appear in the energy momentum tensor. Varying the space-time metric
$G$ in Eq.~\eqcurved, including the implicit dependence made explicit
in Eq.~\eqd, yields a canonical deformation.
\item{2)} (1,1) primary fields are in natural correspondence with the
physical states of string theory, being the vertex operators which
create asymptotic physical states and describe their scattering. This
means that canonical deformations have a straightforward interpretation
as changes in space-time fields.
\item{3)} Canonical deformations work for massive states just as well
as they do for massless, and avoid certain ambiguities and pathologies
which may be implicit in other approaches.

\noindent Note that the third of these virtues seems to involve some
small revision of the standard lore on the relationship between sigma
models and strings: in particular, the, ``standard,'' sigma model,
containing only terms of naive dimension two, not only puts the massive
fields on shell, but also puts them equal to zero.

Appealing though they are, canonical deformations have a significant
drawback; they correspond to turning on space-time fields in a
particular gauge. This is most easily seen in an example. For simplicity,
consider a conformal field theory of free scalars, defined by the energy
momentum tensor of Eq.~\eqcurved\ with $G$ the standard flat Minkowski
metric. A short calculation soon shows that primary (1,1) fields of
naive dimension two are of the form
$$
\Phi_{(1,1)}=H^{\mu\nu}(X)\dx_\mu\dbx_\nu\numbereq
$$
where the coefficient functions $H$ must satisfy certain conditions,
if we are working in the quantum case where all fields are understood
to be normal ordered and the bracket is a commutator
$$\displaylines{\hfill\Box H^{\mu\nu}(X)=0\hfill\llap{\global\advance
\eqnumber by 1 (\the\eqnumber)}\name{\eqeom}\cr
\hfill\partial_\mu H^{\mu\nu}(X)=\partial_\nu H^{\mu\nu}(X)=0\hfill
\llap{\global\advance\eqnumber by 1 (\the\eqnumber)}\name{\eqgauge}\cr}
$$
The first of these is an equation of motion, something that we would expect to
arise in making a conformal deformation, but Eq.~\eqgauge\ is a gauge
condition (of course, (\eqeom) is only the correct equation of motion
when this gauge condition holds). As we shall explain in the next section,
this gauge condition is a serious
nuisance when we come to the problem of symmetry, and motivated us to
explore \ref\gcovdef{M. Evans and I. Giannakis, \it Rockefeller Preprint
\rm RU91-2 (1991)} more general solutions of the deformation equations
(see an alternative approach in \ref\philcov{P.\tie Nelson,
\prl62 (1989), 993; H-S.\tie La and P.\tie Nelson, \np332 (1990), 83.}).

\bigbreak
\line{\bf 4. Symmetries. \hfil}
\nobreak\bigskip

A symmetry is a change in the space-time fields which does not change
the physics ({\it i.e.\/} all masses and S-matrix elements are unchanged).
Since the physics is determined by the conformal field theory corresponding
to the field configuration in question, a transformation on the fields
will be a symmetry if the corresponding conformal field theories are
isomorphic. From our definition of a conformal field theory in
section~2, it is clear what we mean by such an isomorphism; there must
exist an isomorphism of the two operator algebras,
$\rho\colon \cal A_{\rm 1} \rightarrow A_{\rm 2}$ which maps energy momentum
tensors on to one another. In particular, if $\rho$ is an {\it automorphism\/}
such that $\rho(T_\Phi\s)=T_{\Phi+\delta\Phi}\s$, then $\Phi \rightarrow
\Phi+\delta\Phi$ is a symmetry transformation. Here $\Phi$ is a space-time
field configuration, and as such indexes the energy momentum
tensors of conformal field theories; for example, in Eq.~\eqcurved, all
space-time fields $\Phi$ are zero except for the metric $G$.

Even inner automorphisms are interesting in this context, and appear to
give rise to the gauge symmetries of string theory. In this talk we
shall restrict ourselves to inner automorphisms, and so we are
interested in infinitesimal operators $h\in\cal A$ such that
$$
i[h,T_\Phi\s] = T_{\Phi+\delta\Phi}\s - T_\Phi\s \numbereq\name\eqh
$$
(Changing all operators by their commutator with a fixed infinitesimal
operator, $h$, is an algebra isomorphism by virtue of the Jacobi identity;
this is the infinitesimal version of a similarity transformation).
The right hand side of Eq.~\eqh\ is a deformation, which is one reason
for being interested in the subject. There is a large class of operators
$h$ which make the right hand side of Eq.~\eqh\ a canonical deformation:
\smallskip{\narrower\noindent
\sl Let $h$ be the sum of zero modes of (1,0) and (0,1) primary fields,
then $h$ generates a canonical deformation.
\smallskip}\noindent
The proof of this statement [\dcftss] is a straightforward application
of the definition of a primary field, Eq.~\eqprim, and the Jacobi identity.

Despite its simplicity, this is a very interesting class of symmetries.
It includes the familiar general coordinate and two-form gauge invariances,
generated by the, ``obvious,'' such currents $\xi^\mu(X)\dx_\mu$ and
$\zeta^\mu(X)\dbx_\mu$, as well as an infinite class of higher symmetries
generated by currents which classically would have higher dimension, such
as $\psi^{\mu\nu\lambda}(X)\dx_\mu\dx_\nu\dbx_\lambda$. In each case,
these currents are primary fields of dimension $(1,0)$ or $(0,1)$ only
if the {\it parameters\/} of the transformations, $\xi$, $\zeta$,
$\psi \ldots$, satisfy certain differential constraints.

Differential constraints on parameters are quite familiar from field
theory. They are {\it nothing\/} to do with the equations of motion
for the fields, but arise from the demand that a transformation preserve
a {\it gauge condition}. Since canonical deformations are associated with
with gauge conditions, as in Eq.~\eqgauge, it is inevitable that gauge
transformations corresponding to canonical deformations have differential
constraints on their parameters. Correspondingly, if we wish to lift
these constraints we must find a more general set of deformations,
unaccompanied by such gauge conditions.

Exhibiting the explicit transformations on the space-time fields is
possible only for those conformal field theories over which we have
good computational control, such as free bosons. Nevertheless, with
such examples as guides, it is possible to draw certain conclusions
about this class of symmetries [\dcftss]. They are all gauge symmetries,
and most or all of the states of the string are the corresponding gauge
fields. The higher symmetries are spontaneously broken (so that the massive
states are so by virtue of the Higgs mechanism), and mix states at
different mass levels. It is
also possible to argue that the symmetric solutions of string theory
should correspond to topological world-sheet field theories, as
suggested by Witten.

However, to exhibit the claims of the previous paragraph more concretely,
and to use these insights about symmetry to do physics, we need more.
We must understand precisely what the set of generators is for
a wider class of conformal field
theories, and we must relax the differential constraints on the parameters
of the transformations, alluded to above (for the higher symmetries, these
constraints exclude the global transformations which encode so much of
the physics). We shall address the second of these problems and, in so
doing, learn something about the first [\gcovdef]. As we argued above,
if we are to relax the differential constraints, we must first understand
more than canonical deformations, and that will be the subject of the
next section.

\bigbreak
\line{\bf 5. Beyond Canonical Deformations. \hfil}
\nobreak\bigskip

Can we find fully gauge covariant deformations, or does conformal invariance
necessarily come accompanied with a gauge condition? We shall consider
the simplest possible case [\gcovdef], that of turning on massless fields
of
the bosonic string about flat twenty-six dimensional space-time. Thus our
starting conformal field theory is twenty-six free bosons, defined by
Eq.~\eqcurved, with $G$ the canonical Minkowski metric. We shall attempt
to solve the deformation equations, Eq.~\eqdef, but will take a more
general {\it ans\"atz\/} than usual, the most general operator of naive
dimension two:
$$
\eqalign{\delta T&=H^{\nu\lambda}(X)\dx_\nu\dbx_\lambda+
A^{\nu\lambda}(X)\dx_\nu\dx_\lambda\cr
&\hphantom{=H^{\nu\lambda}(X)\dx_\nu\dbx_\lambda}\quad
+B^{\nu\lambda}(X)\dbx_\nu\dbx_\lambda
+C^\nu(X)\dsx_\nu+D^\lambda(X)\dbsx_\lambda\cr
\noalign{\vskip 2\jot}
\delta \Tb&=\ov H^{\nu\lambda}(X)\dbx_\nu\dx_\lambda+
\ov A^{\nu\lambda}(X)\dbx_\nu\dbx_\lambda\cr
&\hphantom{=\ov H^{\nu\lambda}(X)\dbx_\nu\dx_\lambda}\quad
+\ov B^{\nu\lambda}(X)\dx_\nu\dx_\lambda
+\ov C^\nu(X)\dbsx_\nu+\ov D^\lambda(X)\dsx_\lambda\cr}
\numbereq\name{\eqansatz}
$$
The tensors $H^{\nu\lambda}\ldots\ov D^\lambda$ are initially taken
to be completely independent.
This {\it ans\"atz\/} is then substituted into the deformation equations,
Eq.~\eqdef. Some tedious but straightforward manipulation reduces
these conditions, after an appropriate redefinition, to the following:
$$
\eqalign{\delta T\s &= \knl \dx_\nu \dbx_\lambda + \left(\partial
- \ov\partial\right) [C^\nu\dx_\nu - D^\lambda\dbx_\lambda]\cr
\delta \Tb\s &= \knl \dx_\nu \dbx_\lambda - \left(\partial - \ov\partial
\right) [\ov C^\nu\dbx_\nu - \ov D^\lambda\dx_\lambda]\cr}
\numbereq\name{\eqsoln}
$$
Note that $\delta T$ and $\delta \Tb$ differ only by a derivative with
respect to $\sigma$, so that $L_0-\ov L_0$ is invariant, as argued in
section 3.1.
The quantities $C, \ov C, D, \ov D$ are give in terms of $K$, which we
interpret as the sum of the graviton and two-form, by\goodbreak
$$
\eqalignno{\partial_\nu C^\nu&=0&\eqalinno\name{\eqdivc}\cr
D^\lambda&=-\half\partial_\mu K^{\mu\lambda}&\eqalinno\name{\eqd}\cr
\ov D^\lambda&=-\half\partial_\mu K^{\lambda\mu}
&\eqalinno\name{\eqdb}\cr
\partial^\lambda C^\nu&=\half\Box\knl
-\half\partial^\nu\partial_\mu K^{\mu\lambda}&\eqalinno\name{\eqc}\cr
\partial^\lambda\ov C^\nu&=\half\Box K^{\lambda\nu}
-\half\partial^\nu\partial_\mu K^{\lambda\mu}&\eqalinno\name{\eqcb}\cr}
$$

At first sight there is no equation of motion for the physical field
$K$, and the dilaton is nowhere to be seen. However Eqs.~\eqc\ and
\eqcb\ cannot be solved for $C$ and $\ov C$ for arbitrary $K$. There
is an integrability condition which $K$ must satisfy which turns out
to yield both an equation of motion and the dilaton:
$$
\Box\knl-\partial^\nu\partial_\mu K^{\mu\lambda}
-\partial^\lambda\partial_\mu K^{\nu\mu}
+\partial^\nu\partial^\lambda K^\mu{}_\mu
=\partial^\nu\partial^\lambda\phi+\alpha^{\nu\lambda}\numbereq\name\eqeom
$$
for some scalar function $\phi$, which we identify as the dilaton.
Eq.~\eqdivc\ yields the dilaton equation of motion, $\Box\phi=0$.

We have thus achieved our goal of finding deformations which are associated
with a covariant equation of motion and no gauge condition. We get
exactly the expected linearised equations, {\it except\/} for the last
term on the right hand side of Eq.~\eqeom. $\alpha$ is antisymmetric
in its indices and constant. This term therefore indicates
that there is a finite dimensional space of additional deformations,
over and above those associated with turning on the physical fields.

What are we to make of these additional deformations? Some caution
is appropriate [\gcovdef], since our calculations are only to first
order in the deformation, but
there is nonetheless a natural interpretation of these
parameters $\alpha$. With lowered indices, $\alpha$
is a harmonic two-form, and so may be integrated over compact two-dimensional
submanifolds of space-time. This integration may be pulled back to the
world-sheet to yield
$$
S_\alpha = \int  \alpha_{\mu\nu} dX^\mu dX^\nu
\numbereq\name{\eqtop}
$$
which, since $\alpha$ is closed, is invariant under deformations of
the map $X$.
In the language of non-linear sigma
models, $S_\alpha$ is proportional to an instanton number, and when
added to the usual action (a ``$\theta$-term'') will usually have physical
consequences.

In particular, if $S_\alpha$ is added to the usual action with a
coefficient proportional to $\ln\epsilon$ (where $\epsilon$ is, say, the
parameter of dimensional regularisation), then it will clearly amend the
condition for conformal invariance \ref\tseyt{A. Tseytlin, private
communication}. The equations of motion will then
be affected in just the way we see in Eq.~\eqeom. It is worthy
of note that, at least in some cases, the topological charge density
does have just such an ultraviolet divergence at one loop \ref\thesis{M.\tie
Evans, \np208 (1982), 122; Ph.D. Thesis, Princeton University (1983).}.

Given this interpretation of these additional deformations, we may quite
reasonably refer to $\alpha$ either as free parameters or topological
degrees of freedom of string theory. Their description as $\theta$-terms
is very similar to that given by other speakers at this conference of
the isolated states found in two-dimensional string theory [\iso].

\bigbreak
\line{\bf 6. Symmetries Redux. \hfil}
\nobreak\bigskip

Finally, let us set $\alpha=0$ and return to the question of symmetries.
With the apparently covariant set of deformations found in the previous
section, we can now find the {\it full\/} set of unbroken gauge
transformations as inner automorphisms. It is straightforward to see
that
$$
h=\ints\left( \xi^\mu(X)\dx_\mu+\zeta^\mu(X)\dbx_\mu\right),
\numbereq\name{\eqgen}
$$
generates deformations of the type given in Eq.~\eqsoln, and that the fields
transform in the conventional way. Note that we have achieved our goal
of eliminating the differential constraints on the parameters
$\xi$ and $\eta$.

These explicit calculations need to be extended in two ways; we would like
to understand the higher symmetries, and we would like to know how the
generators of symmetry deform, if at all, with the conformal field theory.
To do this
let us first understand why the operators of Eq.~\eqgen\ {\it had\/} to
generate a symmetry. The
argument has two parts: i) turning on the space-time fields without
restricting their gauge corresponded to the most general conformal
deformation {\it by world-sheet fields},
and ii)~commuting the free energy-momentum tensor
with the $h$ of Eq.~\eqgen\ produces a deformation of this form
(by conservation of {\it naive\/} dimension).
Hence it has to be possible to pull back the inner automorphism generated
by $h$ to a symmetry transformation on the space-time fields.

It is
very tempting to generalise this argument to the massive fields
and the higher symmetries. That is, we might conjecture that arbitrary
space-time fields are turned on, without imposing any restriction as
to gauge, by considering an ans\"atz which is the obvious
generalisation of Eq.~\eqansatz\ to the appropriate naive dimension,
and demanding that it be a conformal deformation,
satisfying Eq.~\eqdef. Then a deformation of this
form would necessarily be generated by the corresponding dimensional
generalisation of the generator in
Eq.~\eqgen. The problem is that this symmetry appears to be
larger than we
need; working out some examples [\gcovdef] it is easy to see that
$h$ contains many more degrees of freedom than there are gauge
conditions to relax. It therefore seems likely that the
most general conformal deformation of higher naive dimension contains
not just the physical space-time fields, but also unphysical auxiliary
fields, which are pure gauge artifacts. This is very much akin to the
situation which arises in the superspace formulation of supersymmetric
gauge theories, where there exist auxiliary gauge artifacts over
and above those needed to account for the halving of fermionic
degrees of freedom in going on shell. If we are willing to live
with these auxiliary fields, then we have answered the question
of which operators generate the higher symmetries. On the other
hand, if we wish to restrict ourselves to the physical, propagating
fields with non-trivial dynamics (``Wess-Zumino gauge''),
then it remains to be determined
what the appropriate restrictions on the generators are.

What happens as when we consider a theory other than that of free
bosons? In the discussion of section four, the generators were associated
with primary fields, which deform with the conformal field theory and
so are not known explicitly over the whole deformation class. However
the two preceeding paragraphs contain no mention of primary fields. The
only place where a specific property of the free theory was used was
an occasional appeal to conservation of naive dimension, but this was
only a convenience which allowed us to discuss one mass-level at a
time, and may be dispensed with. It would seem, then, that deforming the
set of operators which generate canonical deformations is complicated
because of the need to preserve a gauge condition, rather than to ensure
an interpretation as a transformation on space-time fields.  Space-time
fields (including auxiliaries) are turned on with the most general
solution of the deformation equations in terms of world-sheet fields,
and so symmetries are generated by all operators which commute with the
generator of translations, $L_0-\ov L_0$. Since this operator is an
invariant of the deformation class, so are the symmetry generators.
In summary:

\smallbreak{\narrower\noindent\sl Symmetries are generated
by the centraliser of $L_0-\ov L_0$, and this set is invariant over
the whole deformation class.\smallbreak}

\bigbreak
\line{\bf Acknowledgements.\hfil}
\nobreak\bigskip
M.E. would like to thank Burt Ovrut for collaboration on some of the early
work described here, and S. Catto and A. Rocha for the invitation to talk
at a very stimulating conference.

\immediate\closeout1
\bigbreak
\line{\bf References.\hfil}
\nobreak\medskip\vskip\parskip

\input refs

\vfill\end